\documentclass[aps,pra,reprint,floatfix,superscriptaddress,showpacs,twocolumn]{revtex4-1}
\usepackage{amsmath}
\usepackage[colorlinks=true,citecolor=blue,linkcolor=blue,urlcolor=blue]{hyperref}
\usepackage[usenames,dvipsnames]{xcolor}
\usepackage{longtable}
\usepackage{graphicx}
\usepackage{color}
\usepackage{float}
\usepackage{placeins}
\usepackage{rotating}
\usepackage{epsfig,rotate}

\begin{document}

	\title{Resonance strengths for~\emph{KLL} dielectronic recombination of highly charged\\
	mercury ions and improved empirical $\boldsymbol{Z}$-scaling law}

	\author{Z.~Harman}
	\email{harman@mpi-hd.mpg.de}
	\author{C.~Shah}
	\email{chintan@mpi-hd.mpg.de}
	\affiliation{Max--Planck--Institut f\"{u}r Kernphysik, Saupfercheckweg 1, 69117 Heidelberg, Germany}
	\author{A.~J.~{Gonz\'alez~Mart\'inez}}
	\affiliation{Max--Planck--Institut f\"{u}r Kernphysik, Saupfercheckweg 1, 69117 Heidelberg, Germany}
	\affiliation{Instituto de Instrumentaci\'on para Imagen Molecular (I3M), Universitat Polit\`ecnica de Val\`encia, Camino de Vera s/n, 46022 Valencia, Spain}
	\author{U.~D.~Jentschura}
	\affiliation{Max--Planck--Institut f\"{u}r Kernphysik, Saupfercheckweg 1, 69117 Heidelberg, Germany}
	\affiliation{Department of Physics, Missouri University of Science and Technology, Rolla MO 65409, USA}
	\author{H.~Tawara}
	\author{C.~H.~Keitel}
	\affiliation{Max--Planck--Institut f\"{u}r Kernphysik, Saupfercheckweg 1, 69117 Heidelberg, Germany}
	\author{J.~Ullrich}
	\affiliation{Max--Planck--Institut f\"{u}r Kernphysik, Saupfercheckweg 1, 69117 Heidelberg, Germany}
	\affiliation{Physikalisch-Technische Bundesanstalt, Bundesallee 100, 38116 Braunschweig, Germany}
	\author{J.~R.~Crespo~L\'opez-Urrutia}
	\affiliation{Max--Planck--Institut f\"{u}r Kernphysik, Saupfercheckweg 1, 69117 Heidelberg, Germany}

	\newcommand{\blue}[1]{{\color{blue}{#1}}}
	\newcommand{\red}[1]{{\color{Red}{#1}}}
	\newcommand{\magenta}[1]{{\color{magenta}{#1}}}
	\newcommand{\green}[1]{{\color{OliveGreen}{#1}}}

	\begin{abstract}
		Theoretical and experimental resonance strengths for ~\emph{KLL} dielectronic recombination (DR) into He-, Li-, Be-, and B-like mercury
		ions are presented, based on state-resolved DR x-ray spectra recorded at the Heidelberg electron beam ion trap.
		The DR resonance strengths were experimentally extracted by normalizing them to simultaneously recorded radiative recombination signals.
		The results are compared to {state-of-the-art} atomic calculations that include relativistic electron correlation and
		configuration mixing effects. Combining the present data with other existing ones, we derive an improved semi-empirical $Z$-scaling law
		for DR resonance strength as a function of the atomic number, taking into account higher-order relativistic corrections, which are
		especially relevant for heavy highly charged ions.
	\end{abstract}
	
	\date{\today}
	
	
	\maketitle

	%
	\section{Introduction}\label{sect:intro}
	%
	
	Charge-state changing processes have an essential importance for the dynamics of plasmas.
	The corresponding reaction rates do not have a monotonic dependence on the absolute charge state, but they rather
	display a more pronounced effect characteristic for the isoelectronic sequence in which the processes take place.
	Understanding these processes therefore requires the knowledge of various atomic processes.
	One of the strongest and most important processes is photorecombination of electrons with ions. It can proceed in a direct, non-resonant,
	and a two-step resonant channel. In the process of radiative recombination (RR), a photon is directly emitted by the recombining electron,
	i.e., it is a time-reverse of the photoelectric effect. Alternatively, in a two-step process, an incoming electron excites a bound electron
	during recombination, leading to dielectronic recombination (DR).
	
	Such resonant photorecombination processes involving highly charged ions (HCI) in collisions with energetic electrons are relevant for a number of
	applications. Indeed, resonant mechanisms are highly efficient in either ionizing or recombining ions and hence DR is of paramount importance for the
	understanding of the physics of outer planetary atmospheres, interstellar clouds. It is also a very effective radiative cooling mechanism in
	astrophysical~\cite{Massey42,Burgess64,hitomi2016} and laboratory plasmas~\cite{Cohen90,Cummings90}. Thus, a precise quantitative understanding of such process is indispensable. DR often represents the dominant pathway for populating excited states in plasmas and, consequently, for inducing easily observable
	x-ray lines which are used as a diagnostic tool for fusion plasmas~\cite{Widmann95,beiersdorfer2015}, triggering a range of DR studies with highly
	charged ions~\cite{Bitter93,Fuchs98,Radtke00}. In addition to RR and DR, trielectronic recombination was recently emphasized to be crucial for
	plasma models. Recent experiments have shown that~\emph{intra-shell} trielectronic recombination dominates the recombination rates in
	low-temperature photoionized plasmas~\cite{Schnell03,Orban10}. Also, an \emph{inter-shell} trielectronic recombination channel was measured
	to have sizable and even high cross sections relative to first-order DR for low-\textit{Z} elements~\cite{Beilmann11,Beilmann13,Baumann14,Shah16,Shah18},
	and hence, is crucial for high-temperature collisionally ionized plasmas.
	
	From a more fundamental point of view, the selectivity of DR allows stringently testing sophisticated atomic structure calculations, in particular
	of relativistic and quantum electrodynamics (QED) effects in bound electronic systems. Investigating HCIs with DR offers additional important
	advantages, including large cross sections, the simplification of the theory due to a reduced number of electrons, and pronounced relativistic
	and QED contributions. These have been investigated in experiments both at electron beam ion traps (EBITs) (see, e.g.,
	\cite{Knapp95,Gonzalez05,Gonzalez06,ZH06,Zou03,Nakamura08}) and at storage rings~\cite{Kilgus90,Kilgus93,Man98,Bra02,Bra03,Ma03,Schnell03,Lestinsky08,Bra08,Orban10,Ber15,Ber15JPB}.
	Even if direct EBIT spectroscopic measurements have achieved higher precision~\cite{Bei05}, we can point out that the $2s_{1/2}-2p_{1/2}$ splitting in lithiumlike ions was
	determined in a storage ring employing DR with an accuracy capable of testing two-loop QED corrections~\cite{Bra03}. Similarly, using DR in an ultra-cold electron target,
	the same splitting in Li-like Sc${}^{18+}$ has been indirectly determined with a 4.6-ppm
	precision~\cite{Lestinsky08}. DR experiments have also shown to be sensitive to isotopic shifts in Li-like ${}^{142,150}$Nd~\cite{Bra08,Sch04}.
	
	Early EBIT measurements of DR cross sections and studies at high collision energies, involving quantum interference effects between the RR and
	DR processes in ions up to U${}^{88+}$~\cite{Knapp95} demonstrated the tremendous potential of the method. Previously, we have observed the quantum
	interference phenomenon in a state-specific manner~\cite{Gonzalez05}. We have also succeeded, for the first time, in determining the absolute DR
	resonance energies in HCI in a state-resolved fashion, including He-like mercury ions (Hg$^{78+}$)~\cite{Gonzalez06} with high precision of a few
	eV on a 50~{keV} energy range. These results have been compared to advanced relativistic theoretical calculations, such as the multiconfiguration
	Dirac-Fock (MCDF) method and a configuration interaction scheme employing a combined Dirac-Fock-Sturmian basis set (CI-DFS), both including
	quantum electrodynamic (QED) contributions~\cite{ZH06}. While, generally, a very good agreement between theory and experiment has been observed
	(on the level of a few ppm), some potentially interesting disagreements remain to be addressed.
	
	In addition to such structural investigations, another important features of photorecombination processes are cross sections and strengths.
	Since the resonant excitation in DR is solely evoked by the interaction of the active electrons, the experimental determination of cross sections
	provides one new insights into relativistic electron interactions in a dynamical process. Recently, the experiments became sensitive to the
	contribution of the generalized Breit interaction~\cite{Nakamura08,Bernhardt11} to DR resonance strengths, as well as to the linear polarization
	of x rays emitted during DR~\cite{Fritzsche09,Shah15}. Also, the theoretical description of the
	process requires non-trivial additions to the many-body theory of atomic structures. In our case, the MCDF method is applied to describe the bound
	few-electron states involved in the process, and a relativistic distorted-wave model of the continuum electron is employed.
	
	Several experimental as well as theoretical studies on DR cross sections $\sigma^{\mathrm{DR}}$ and resonance strengths $S^{\mathrm{DR}}$ have
	been performed for intra- as well as inter-shell transitions. A specific example of inter-shell dielectronic excitations are the~\emph{KLL}
	transitions. These take place when a free electron is captured into a vacant state of the \emph{L}-shell of an ion, while a bound electron of
	the ion from the \emph{K}-shell is simultaneously promoted to the \emph{L}-shell, thus forming an intermediate autoionizing $1s2l2l'$ state.
	So far many experimental investigations have been reported on~\emph{KLL} DR resonances of various low- and mid-$Z$
	ions~\cite{Kilgus93,Ali11,Ali91,ORourke04,Beiersd92,Knapp89,Zhang04,Fuchs98,Knapp93,Watanabe07,Yao10}, while data are rather scarce for very
	heavy ions where relativistic and QED effects play a critical role~\cite{Kavanagh2010,Tu2016}, and therefore a full scope has been still missing.
	
	In the present paper, we investigate and determine state-resolved \emph{KLL} DR resonance strengths for highly charged mercury ions in different
	charge states (Hg$^{78+}$ to Hg$^{75+}$) using the Heidelberg EBIT and compare them to calculations based on the MCDF method, and the Flexible
	Atomic Code (FAC). Experimental DR spectra are normalized to the radiative recombination cross section in order to obtain the resonance strengths.
	In Section~\ref{sect:theory}, the theoretical calculations are briefly described. The experimental procedure and data analysis are described in
	Section~\ref{sect:exp}, and theoretical and experimental results are compared. Then, in Section~\ref{sect:scaling}, combining the experimental
	results available so far, including the new data for Hg ions in the present work, we provide a new semi-empirical formula to describe~\emph{KLL} DR
	strengths for He-like ions over a wide range of nuclear charges. The paper concludes with a Summary (Section \ref{sect:summary}). Atomic units
	are used ($\hbar=m_e=e=1$), unless noted otherwise.
	
	%
	%
	\section{Theory and calculation of resonance strengths}\label{sect:theory}
	%
	%
	
	The cross section for a given dielectronic recombination channel is given (in atomic units) as a function of the electron kinetic energy 
	$E$ as (see, e.g.~\cite{HaanJacobs,Zim90,Zimmermann})
	\begin{equation}
	\label{eq:drkompakt}
	\sigma^{{\mathrm{DR}}}_{i \to d \to f}(E) =
	\frac{2\pi^2}{p^2}  V_a^{i\to d} \frac{A_r^{d \to f}}{\Gamma_d}  L_d(E)ű,.
	\end{equation}
	The Lorentzian line shape function
	\begin{equation} 
	L_d(E) =  \frac{\Gamma_d/(2\pi)}
	{(E_i+E-E_d)^2 +\frac{\Gamma_d^2}{4}}
	\end{equation}
	is normalized to unity on the energy scale and $p=|\vec{p}|= \sqrt{(E/c)^2 - c^2}$ is the modulus of the free-electron momentum associated with the kinetic energy $E$. Furthermore, $\Gamma_d$ denotes the total natural width of the intermediate autoionizing state, given as the sum of the radiative and autoionization widths: $\Gamma_d = A^d_r + A^d_a$ (note that rates and the associated line widths are equivalent in atomic units). 
	In Eq.~\eqref{eq:drkompakt}, $i$ is the initial state of the process, consisting of the ground-state ion and a continuum electron with an asymptotic momentum $\vec{p}$ and spin projection $m_s$. The wave function of the latter is represented by a partial wave expansion~\cite{Eichler},
	\begin{eqnarray}
	|E \vec{p} m_s\rangle&=&\sum_{\kappa m}i^l e^{i\Delta_{\kappa}} 
	\sum_{m_l}Y_{l m_l}^*(\theta,\varphi) \\
	&\times& C\left(l\ \frac{1}{2}\ j;m_l\ m_s \ m\right)| E \kappa m\rangle \,, \nonumber
	\end{eqnarray}
	where the orbital angular momentum of the potential wave is denoted by $l$ and the corresponding magnetic quantum number is $m_l$. The phases $\Delta_{\kappa}$ are chosen so that the continuum wave function fulfills the boundary conditions of an incoming plane wave and an 
	outgoing spherical wave, as necessary for the description of an incoming electron ({\em sic}, see Ref.~\cite{Eichler}).
	In the above expression, $\kappa = 2(l-j)\,(j+1/2)$ is the relativistic angular momentum quantum number. The total angular momentum quantum number of the partial wave $| E \kappa m\rangle$ is $j=|\kappa|-\frac{1}{2}$. 
	The spherical angular coordinates are denoted by $\theta$ and $\varphi$, $Y_{lm_l}(\theta,\varphi)$ is a spherical harmonic and the $C\left(l\ \frac{1}{2}\ j;m_l\ m_s\ m\right)$ stand for the vector coupling coefficients.
	The partial wave functions are represented in the spherical bispinor form as
	\begin{equation}
	\langle \vec{r}| E \kappa m \rangle =
	\psi_{E \kappa m}(\vec{r})= \frac{1}{r}
	\left(\begin{array}{c} P_{E \kappa}(r)\Omega_{\kappa m}(\theta,\varphi)\\
	i Q_{E \kappa}(r)\Omega_{-\kappa m}(\theta,\varphi)\end{array}
	\right)\ .
	\end{equation}
	Here, $P_{E \kappa}(r)$ and $Q_{E \kappa}(r)$ are the radial parts of the large and small component wave functions, and $\Omega_{\kappa m}(\theta,\varphi)$ is the spinor spherical harmonic in the $lsj$ coupling scheme.
	
	The index $d$ in Eq.~(\ref{eq:drkompakt}) denotes quantities related to the autoionizing state formed which constitutes the intermediate state in the dielectronic capture process. This intermediate state then decays radiatively to the final state $f$. $V_a^{i \to d}$ denotes the dielectronic capture (DC) rate and $A_r^{d}=\sum_f A_r^{d \to f}$ is the total radiative rate of the autoionizing intermediate state $|d\rangle$. The DC rate is given by
	\begin{eqnarray}
	\label{eq:dr-rate}
	V_a^{i \to d} &=& \frac{2\pi}{2(2J_i+1)} \sum_{M_{d}} \sum_{M_i m_s}
	\int \sin(\theta) d\theta d\varphi \\
	& & |\langle\Psi_{d}; J_{d} M_{d}  | V_C + V_B | \Psi_i E; J_i M_i, \vec{p}m_s\rangle|^2 \nonumber \\
	&=& 2\pi \sum_{\kappa} |\langle\Psi_{d}; J_{d} || V_C + V_B 
	|| \Psi_i E; J_i  j; J_d \rangle|^2 \,. \nonumber
	\end{eqnarray}
	In this equation, the matrix element of the Coulomb and Breit interaction~\cite{Breit29} ($V^C$ and $V^B$, respectively) is calculated for the initial bound-free product state $i$ and the resonant intermediate state $d$. After integration over the initial magnetic quantum numbers and the direction $(\theta,\varphi)$ of the incoming continuum electron, and after performing the summation over the magnetic quantum numbers of the autoionizing state, we obtain the partial wave expansion of the reduced matrix elements, as given in the last line of the above equation.
	
	The dielectronic capture rate is related to the rate of its time-reversed process, i.e., the Auger process, by the principle of 
	detailed balance:
	\begin{equation} 
	\label{balance}
	V_a^{i \to d} = \frac{2J_{d}+1}{2(2J_i+1)} A_a^{i \to d} \,.
	\end{equation}
	Here, $J_d$ and $J_i$ are the total angular momenta of the intermediate and the initial states of the recombination process,
	respectively. Neglecting the energy-dependence of the electron momentum in the vicinity of the resonance, the dielectronic resonance
	strength, defined as the integrated cross section for a given resonance peak,
	\begin{equation} 
	S^{\mathrm{DR}}_{i \to d \to f} \equiv \int \sigma^{{\mathrm{DR}}}_{i \to d \to f}(E) dE\,,
	\end{equation}
	is given as
	\begin{equation} 
	S^{\mathrm{DR}}_{i \to d \to f}
	= \frac{2\pi^2}{p^2} \frac{1}{2} \frac{2J_{d}+1}{2J_i+1} \frac{A_a^{i \to d}A_r^{d \to f}}{A_r^{d}+A_a^{d}}\,,
	\label{eq1}
	\end{equation}
	where $A_a^{i \to d}$ is implicitly defined in Eq.~\eqref{balance}. The factor $\frac{2\pi^2}{p^2}$ defines the phase space density and the $1/2$
	stems from the spin degeneracy of the free electron. 
	
	To obtain the cross section corresponding to a given photon emission polar angle $\theta$, the differential cross section for dipole x-ray emission has to be determined.
	{For electric dipole transitions relevant to the current study, it is given by~\cite{ChenScofield}}
	\begin{eqnarray}\label{eq:ang}
	\frac{d\sigma^{\mathrm{DR}}_{i\to d \to f}}{d\Omega_k} &=&
	\frac{\sigma_{i\to d \to f}^{\mathrm{DR}}}{4\,\pi} \,\,\, W(\theta) \,,\\
	W(\theta)&=&
	\left(1+\beta_{i \to d \to f} P_2(\cos \theta)\right) \,. \nonumber
	\end{eqnarray}
	Also, the resonance strength has to be modified accordingly, i.e.~multiplied by the angular distribution function $W(\theta)$. In the
	above formula, $\beta_{i \to d \to f}$ is the dipole anisotropy parameter depending on the matrix elements of dielectronic capture and
	on the angular momentum quantum numbers of the initial and intermediate states involved in the electron recombination and $P_2(x)$ is the
	second-order Legendre polynomial. The anisotropy parameter can be expressed as~\cite{ChenScofield,Gai98} (see also \cite{Zak03JPB,Zak03})
	\begin{equation}\label{eq:beta}
	\beta_{i \to d \to f} \! = \! 
	\frac{(-1)^{1+J_d+J_f} P_{J_iJ_d}^{(2)}}{P_{J_iJ_d}^{(0)}}
	\sqrt{ \tfrac{3}{2}(2J_d+1)}  \!
	\left \{\begin{array}{lcl} 1 & 1 & 2 \\
	J_d & J_d & J_f \end{array} \right \}
	\end{equation}
	with
	\begin{eqnarray}
	\label{pjijd2}
	&& P_{J_iJ_d}^{(L)} = \sum_{\kappa \kappa'}(-1)^{J_i+J_d+L-1/2} i^{l-l'}
	\cos(\Delta_{\kappa} - \Delta_{\kappa'}) \\
	&&      \times [j,j',l,l',L]^{1 \over 2}
	\left (\begin{array}{lcl}
	l & l' & L \\
	0 & 0 & 0
	\end{array} \right ) 
	\left \{\begin{array}{lcl}
	j' & j & L \\
	l & l' & {1 \over 2}
	\end{array} \right \} 
	\left \{\begin{array}{ccl}
	J_d & J_d & L \\
	j & j' & J_i
	\end{array} \right \} \nonumber \\ 
	&&     \times  \langle\Psi_d; J_d \| V_C + V_B \| \Psi_i E; J_i j ;J_d\rangle \nonumber \\
	&&     \times  \langle\Psi_d; J_d \| V_C + V_B \| \Psi_i E; J_i j';J_d\rangle^* \,.  \nonumber
	\end{eqnarray}
	Here, the shorthand notation $[j_1,j_2,\dots,j_n]=(2j_1+1)(2j_2+1)\dots(2j_n+1)$ is used.  We denote $3j$ symbols with round brackets and represent $6j$ symbols by curly brackets.
	
	In this work, we observed the x-ray radiation at $\mathrm{90}^\circ$ to the electron beam propagation direction. 
	Thus, according to Eq.~\eqref{eq:ang}, the angular correction factor for electric dipole x-ray transitions can be given as,
	\begin{equation}
	W(\mathrm{90}^\circ) = \frac{3}{3 - P^{\mathrm{DR}}},
	\end{equation}
	where $P^{\mathrm{DR}}$ is linear polarization of DR x rays.

	%
	\section{Experimental resonance strengths}
	\label{sect:exp}
	%
	%
	%
	\subsection{Experiment and data analysis}
	%
	%
	
	\begin{figure}
		\begin{center}
		\includegraphics[width=0.95\columnwidth]{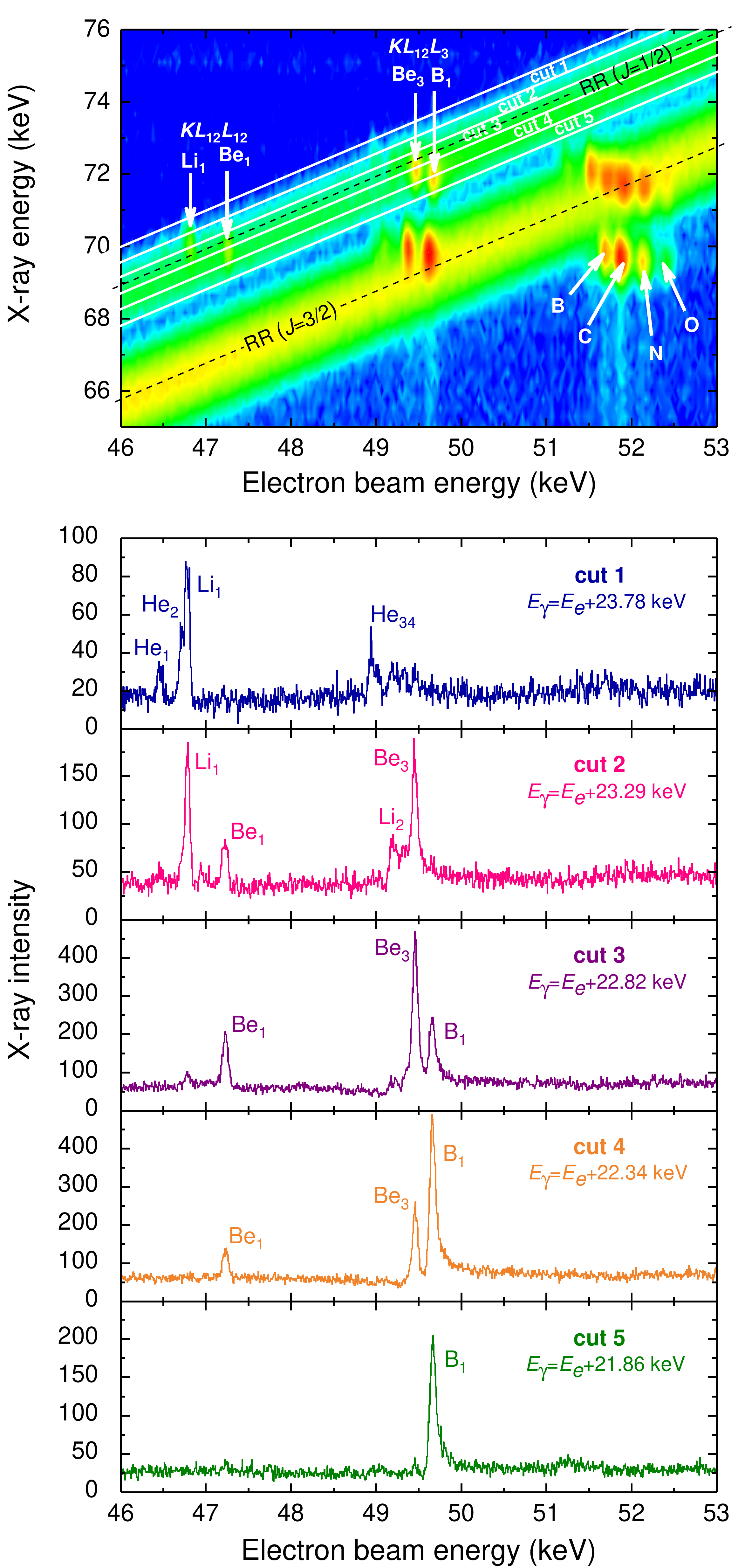}
		\caption{(Color online) Upper panel: A typical 2D plot of the observed ~\emph{KLL} DR and RR x rays from Hg ions in different charge states
		as a function of the electron beam energy. The element symbol refers to the initial charge state of the Hg ions.
		Lower panel: An example of projections of the sliced portions in the $J=1/2$ region at different RR x-ray energies, along the electron
		energy axis. Cut 1 corresponds to a slice at the highest RR x-ray energy. The background is due to RR, and the observed peaks are due
		to \emph{KLL} DR of Hg ions in different initial charge states as indicated with He-, Li-, Be-, and B-like Hg ion. See the text for
		further detailed explanations.}\label{scatter}
		\end{center}
	\end{figure}
	
	The present experiment with highly charged mercury ions (He- to B-like) was carried out using the HD-EBIT~\cite{Cres99} at the Max~Planck~Institute for Nuclear Physics in Heidelberg. 
	Experimental details have already been discussed in previous papers~\cite{Zhang04,Gonzalez05,Gonzalez06}. 
	It should be pointed out that relative resonance energies were precisely determined with uncertainties of approximately 4~eV at a 50~keV DR resonance region, corresponding to a resolution of $\Delta E/E \approx 10^{-4}$, while the electron beam energy spread was estimated to be about 60~eV FWHM at 50~keV.

	We generate two-dimensional (2D) plots displaying the x-ray energy against the electron beam energy
	which is slowly scanned over the region of \emph{KLL} DR resonances. The top panel of Fig.~\ref{scatter} shows a typical 2D plot of such scans for Hg ions
	including different charges, with an acquisition time of about 100 hours. For a given charge state and capture level, the energy scan
	register a unity-slope band, broadened both by the energy spread of the electron beam and the energy resolution of the photon detector.
	The two broad bands in Fig.~\ref{scatter} (top panel) correspond to the RR into $n=2$ states with different total angular momenta $J$ of the final,
	bound many-electron state: the one at higher x-ray energy (lower electron beam energy) is due to RR into the $n=2$ state with $J = 1/2$,
	meanwhile the other band at lower x-ray energy is due to $n=2$, $J=3/2$ states. A number of bright spots---DR resonances---appear at
	specific electron and photon energies. They are mostly overlapping with the RR broad bands and are observed to cluster around three
	energy regions such as $KL_{12}L_{12}$, $KL_{12}L_3$, and $KL_{3}L_3$. These resonances correspond to different ionic states involved in
	the DR process. For example, $KL_{12}L_{12}$ represents~\emph{KLL} DR with both the initially free electron as well as a
	\textit{K}-shell electron being promoted into an $n=2$, $J=1/2$ state, forming either a $1s 2s_{1/2}^2$, $1s 2s_{1/2}2p_{1/2}$ or a
	$1s 2p_{1/2}^2$ intermediate excited configuration state.

	The data on the 2D plot can be sliced and projected onto either the electron beam energy or x-ray energy axis.
	In fact, the projection into the electron beam energy axis of thin portions sliced along the RR band (at either $J=1/2$ or $J=3/2$) in this 2D plot allows us to investigate the detailed properties of the DR resonances for a given charge state~\cite{Gonzalez05,Gonzalez06}.
	In the bottom panel of Fig.~\ref{scatter}, we demonstrate how we have sliced this plot into relatively narrow widths (white lines), separating the contribution to the DR resonances of Hg ions in different ionic charge states
	and electronic states: namely, the sliced band at the highest x-ray energy (marked as cut~1) mainly consists of those from He-like and Li-like ions. 
	The former are hardly seen in the upper panel of Fig.~\ref{scatter} but are clearly seen in the projections of the lower panel.
	Some examples sliced into narrow widths ($\approx$~500 eV) along different RR x-ray energies and projected onto the electron energy axis are shown in the
	lower panel of Fig.~\ref{scatter}, where one can see a number of peaks corresponding to DR resonances of Hg ions in different initial charge and ionic
	states.	In the top figure sliced at the highest RR x-ray energy region (cut 1), we can clearly see the DR resonances of He-like ions (one into $KL_{12}L_{12}$, marked as He$_1$ and another into a $KL_{12}L_3$ state, He$_3$) and Li-like ions (into $KL_{12}L_{12}$, Li$_1$) at different electron energies. 
	On the other hand, cut 5 at the lowest x-ray energy is dominated by the contribution of $KL_{12}L_3$ DR into B-like ions (marked as B$_1$). 
	The labeling of these resonances has been described in Refs.~\cite{Gonzalez06,ZH06}.
	
	Most experiments could not separate the DR into different states due to limited energy resolutions, their DR strengths should be considered as values summed
	over the possible DR resonances within a certain manifold of atomic states~\cite{Knapp89,Knapp93,Watanabe01}.
	Because of the good electron beam energy resolution and a relatively large separation among different electronic states of heavy Hg ions in the present experiment,
	we can determine experimental resonance strengths of each DR resonances by integrating the counts under the observed DR peak {shown in the lower panel} of Fig.~\ref{scatter}.
	However, determining the absolute resonance strengths requires the knowledge of the number of ions in the trap and the overlap between the electron beam and ion cloud.
	Since DR and RR occur in the same ion-electron collision volume in the present EBIT experiment and RR rates are proportional to the ion number density and overlap factors,
	it is most convenient to normalize the observed DR x-ray intensities to the RR x-ray intensity to determine the absolute resonance strengths.
	Moreover, the RR cross sections ($\sigma^{\mathrm{RR}}$) can be calculated very accurately when the electron beam energy is high, as in our case.
	The theoretical RR cross sections are also less susceptible to correlation effects. Therefore, using the method used by Smith~\textit{et al.}~\cite{Smith00}, we can write:
	\begin{equation}\label{normalize}
	S^{\mathrm{DR}} = \frac{I^{\mathrm{DR}} (3 - P^{\mathrm{DR}})}{I^{\mathrm{RR}} (3 - P^{\mathrm{RR}})} \, \sigma^{\mathrm{RR}} \, \Delta E \, 4 \pi \,, 
	\end{equation}
	where $I^{\mathrm{DR}}$ is the x-ray intensity integrated under a particular~\emph{KLL} DR resonance peak, observed at 90 degrees in the present work, and $I^{\mathrm{RR}}$ is the
	integrated intensity of the RR contribution in the range of the DR peak that has a width of $\Delta E$.
	Since the ions in the EBIT are excited by a unidirectional electron beam, the x-ray photons emitted from the trap are usually anisotropic and polarized~\cite{ChenScofield,Shah15,Shah18}.
	The factors $P^{\mathrm{DR}}$ and $P^{\mathrm{RR}}$ are the polarization factors of x rays emitted from the ~\emph{KLL} DR and the RR processes, respectively, given as
	$P=3\beta/(\beta-2)$ in terms of the electric dipole anisotropy parameter $\beta$ (see Eqs. \eqref{eq:beta} and \eqref{eq:ang}). 
	The factor $4 \pi$ converts differential cross sections for emission at $\mathrm{90}^\circ$ to the electron beam to the total cross sections.
	
	It is important to note that a significant distortion of the continuous and smooth RR x-ray backgrounds ($I^{\mathrm{RR}}$) can be caused by quantum mechanical interference between
	the DR and RR pathways which becomes significant for very heavy ions~\cite{Gonzalez05}. To avoid such effects, we have taken $I^{\mathrm{RR}}$ at slightly below and above the beam energies
	at which DR resonances occur, and used their average in the analysis of Eq.~\eqref{normalize} instead of those directly under the DR resonance peak.
		
	In the present experiment, the ion charge in the EBIT is not well defined but it is distributed over a range of possible charge states of the ions;
	as an example, He- to F-like Hg ions can contribute to the present RR bands into $n=2$ states.
	Therefore, we need to accurately know the \textit{relative} fractional distributions of ions in different charge states to obtain the DR strength for a particular charge state
	as the observed RR x rays ($I^{\mathrm{RR}}$) are the sum of those from all of the possible ions with different charges.
	
	\begin{figure}
		\begin{center}
			\includegraphics[width=0.95\columnwidth]{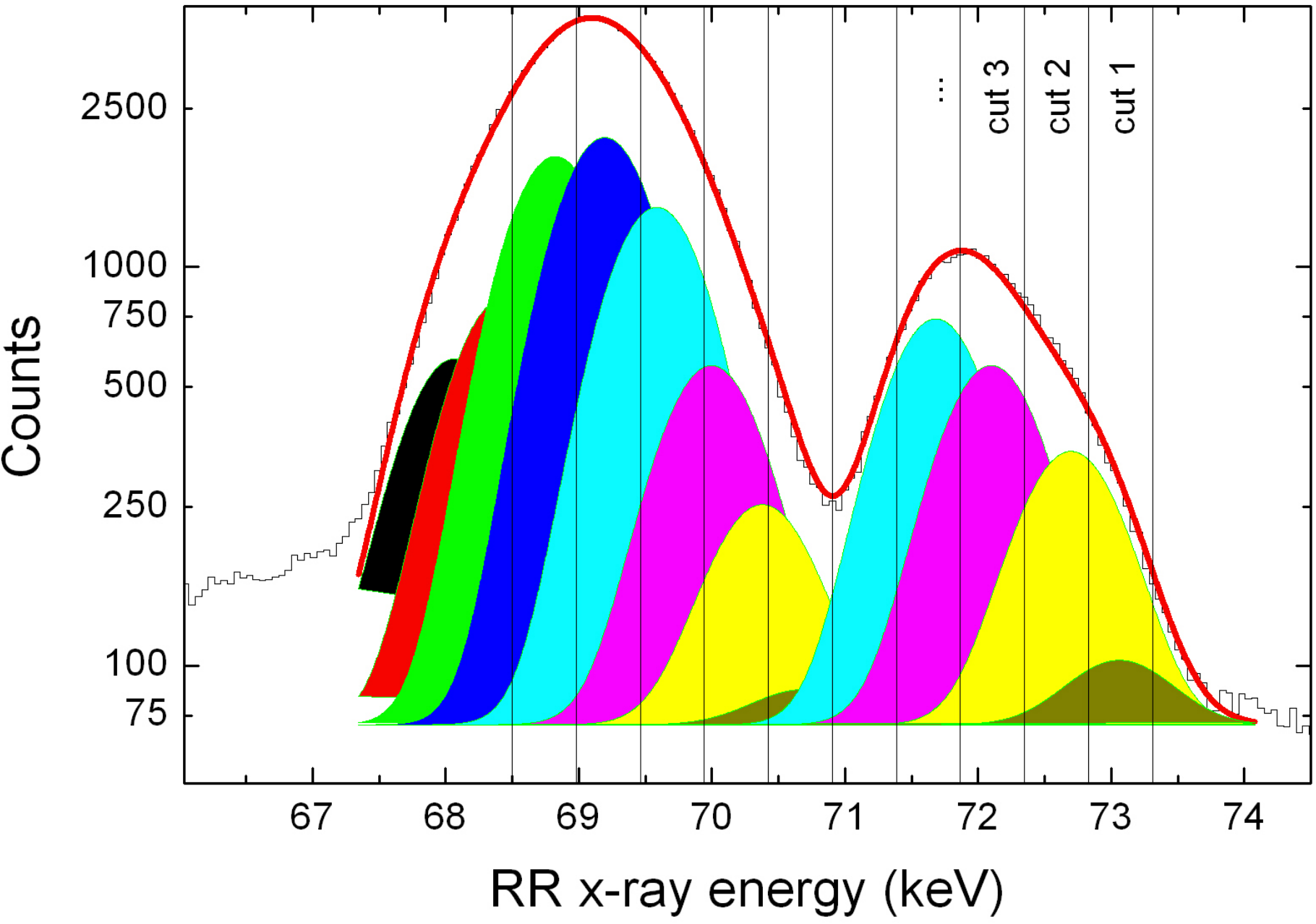}
		\end{center}
		\caption{
			(Color online) Fractional distribution of Hg ions in different charge states contributing to two RR bands ($J=3/2$, on the left-hand side, and $J=1/2$, on the right-hand side). 
			Note that the RR band with $J=1/2$ consists of four charge states, while that with $J=3/2$ consists of eight charge states. 
			The vertical thin lines show the cuts corresponding to the cuts in Fig.~\ref{scatter}. 
			The brown-colored area corresponds to RR into He-like ions, yellow: Li-like ions; red: Be-like ions; green: B-like ions; blue: C-like ions; light green: N-like ions; magenta: O-like ions and dark blue: F-like ions.
		}
		\label{RRbands}
	\end{figure}

	To obtain information on the charge fraction distributions of Hg ions in the trap, we have used the diagonal RR bands. 
	We then selected four electron energy regions (well outside the DR resonances to avoid any distortion effect of the RR spectrum) after sliced vertically and projected the summed
	spectrum onto the x-ray axis. The final profile has been found to contain two strong bumps as shown in Fig.~\ref{RRbands}, where a peak at higher energy corresponds to the RR $J=1/2$ band,
	while a broader peak at lower energy to the RR $J=3/2$ band. The peak observed at higher RR x-ray energy is composed of four sub-peaks, corresponding to RR into the four possible vacancies
	in the $2s_{1/2}$ and $2p_{1/2}$ states with $J=1/2$ in He-, Li-, Be-, and B-like ions. Because the observed RR spectrum depends on RR cross sections and on the number of ions
	in different charge states present in the EBIT, we can estimate the fractional charge distribution of the ions contributing to RR via an analysis of the RR spectrum distributions.
	
	In the present analysis of the RR band spectrum at higher energies (recombination into $J=1/2$ states), we have first set a single constraint: the difference of the observed
	RR x-ray peak energies among different ion charges is set equal to that of the respective theoretical ionization energies as the RR x-ray energy is linearly varied against the
	ionization energy of ions to be recombined~\cite{Sco03}. Convolving the calculated RR cross sections for each ion charge state with the energy resolution of the detector,
	we could fit the observed RR band reasonably well (on the right-hand side in Fig.~\ref{RRbands}) with these four RR peaks from He- to B-like Hg ions.
	The charge fractions obtained are shown in the first row of Table~\ref{population}. The fraction of He-like ions is indeed very small compared to those of the Be- and B-like ions.
	
	\begin{table}[t]
		\caption{
			Percentages (\%) of Hg ions in various charge states contributing to the two RR bands (the $J = 1/2$ and $J = 3/2$ are in the upper and lower parts, respectively) as well as to x-ray intensities in the corresponding selected cuts. The designation of the cuts corresponds to that in Fig.~\ref{scatter}.
			A large fraction ($\approx 66$\%) in the $J = 3/2$ RR band is due to the relatively lower charge states, i.e., C- to F-like. Note that their fractions are not shown here.}
		\begin{center}
			\begin{tabular}{ccccc}
				\hline\hline
				& He &  Li &  Be &  B  \\
				\hline 
				RR, $n=2$, $J=1/2$   & 1.6 &	17.8	&33.9 &	45.0 \\
				\hline
				cut 1	&13.3	&74.7	&11.2& \\	
				cut 2	&&	45.8	&42.0	&9.5 \\
				cut 3	&&	9.3	&50.7&	39.5 \\
				cut 4	&&&		27.0	&70.0 \\
				cut 5	&&&&			78.1 \\
				\hline
				RR, $n=2$, $J=3/2$ & 0.2 &	2.9	& 8.4 &	22.9 \\
				\hline
				cut 6	&4.5&	40.6&	35.9	&\\
				cut 7	&&	13.7&	38.0	&38.6 \\
				cut 8	&&&		13.4	&46.0 \\
				cut 9	&&&		1.8	&21.0 \\
				\hline\hline
			\end{tabular}
		\end{center}
		\label{population}
	\end{table}

	The second, broader band at lower energies due to RR into $J = 3/2$ states shown in Fig.~\ref{RRbands} originates from RR into ions with eight different charge states ranging
	from He- to F-like because the corresponding x-ray energies lie in a close range. The constraint in fitting the second band was analogous to the one used in the analysis of
	the first band. Additionally, to ensure the relation of both RR into $J = 1/2$ and $J = 3/2$ peaks, two more constraints were set in the present analysis:
	First, all peak widths were set to the x-ray detector resolution $\approx$ 676 eV at 73 keV. Second, the radiative recombination into Be-like has only two possible direct
	electron captures, RR into $J = 1/2$ and $J = 3/2$, yielding B-like ($2p$) Hg. Therefore, the difference between the RR x-ray peak energies into $J = 1/2$ and $J = 3/2$
	bands of Be-like ions was fixed to the theoretically calculated one. The best fitting obtained in the second band ($J = 3/2$) is shown on the left-hand side of Fig.~\ref{RRbands}.
	Thus, we were able to determine the relative fractions of Hg ions in different charge states contributing to the observed RR band with $J = 3/2$ which are summarized in the second
	row in Table~\ref{population}. Roughly $2/3$ of ions in the trap are in lower charge states such as C-like to F-like, which do not contribute to the present data analysis.
	
	Now, we have to find the real fractions of ions in a particular charge state contributing to RR and DR in a series of the present cuts shown in Fig.~\ref{scatter}.
	After we have set the slice lines at the same RR x-ray energies as in Fig.~\ref{scatter}, we estimated the fraction of ions in
	\textit{a particular charge state in a specific cut} through the fitted Gaussian distributions. They are shown in the lower part of Table~\ref{population}.
	Using these fractional distributions of ions in different charge states, we can obtain the DR resonance strengths using Eq.~\eqref{normalize}.
	Using this procedure which combines theoretical analysis of a well-understood process (RR) into ions with different charge states with experimental input from the
	two broad-band structures in Fig.~\ref{scatter}, we could finally normalize the DR resonances to the RR cross sections for each individual DR process.

	%
	\subsection{Comparison with theory}\label{sect:comp}
	%
	%
	
	\begin{table*}[t]
		\caption{\label{theory-exp}
			Comparison of measured and calculated ~\emph{KLL} DR strengths S$^{\mathrm{DR}}$ (in 10$^{-20}$ eV cm$^2$) for different He-, Li-, Be-, and B-like states. 
			The DR resonances with the centroid energies $\mathrm{E}_{\mathrm{res}}^{\mathrm{DR}}$ are labeled by the initial charge states of the recombining ion followed by a number and identified by the autoionizing states. 
			The resonances are given in \textit{j--j} coupling notation, where the subscripts after the round brackets stand for the angular momentum of the coupled sub-shells and those after the square brackets denote the total angular momentum of the state. 
			The theoretical DR strengths S$^{\mathrm{DR}}$, radiative recombination cross sections $\sigma$$^{\mathrm{RR}}$ (in 10$^{-23}$ cm$^2$) are calculated with various atomic codes, MCDF$_{m}$ (this work), MCDF$_{s}$ (by Scofield) and FAC (this work). 
			$P^{\mathrm{DR}}$ and $P^{\mathrm{RR}}$ represent the calculated polarization of x rays emitted in the radiative recombination and dielectronic recombination processes, respectively. 
			The theoretical results are given for the case of the full inter-electronic interaction with the Breit term included, represented by (C+B). 
			Experimental uncertainties are given as $1\sigma$. 
		}
		\begin{tabular}{cllcccccccccc}
			\hline \hline
			Label & Autoionizing State & \multicolumn{2}{c}{Experiment} &       & \multicolumn{8}{c}{Theory} \\
			&       & $\mathrm{E}_{\mathrm{res}}^{\mathrm{DR}} (\mathrm{keV})$ & $\mathrm{S}^{\mathrm{DR}}$ & \multicolumn{3}{c}{$\mathrm{S}^{\mathrm{DR}}$ (C+B)} & \multicolumn{2}{c}{$\mathrm{P}^{\mathrm{DR}}$ (C+B)} & \multicolumn{2}{c}{$\mathrm{\sigma}^{\mathrm{RR}}$} & \multicolumn{2}{c}{$\mathrm{P}^{\mathrm{RR}}$} \\
			&       &   &  & MCDF$_m$ & MCDF$_s$ & FAC   & MCDF$_m$ & FAC   & MCDF$_s$ & FAC   & MCDF$_s$ & FAC \\
			\hline \hline
			He$_1$ & $[1s (2s^2)_{0}]_{1/2}$ & 46.358(4) & 3.61 $\pm$ 0.72 & 3.16  & 3.16  & 3.49  & 0.00  & 0.00  & 5.43  & 4.96  & 0.87  & 0.88 \\
			He$_2$ & $[(1s 2s)_{0} 2p_{1/2}]_{1/2}$ & 46.611(6) & 6.30 $\pm$ 0.97 & 4.86  & 4.97  & 5.39  & 0.00  & 0.00  & 5.39  & 4.92  & 0.87  & 0.88 \\
			He$_{34}$ & $[(1s 2s)_{0} 2p_{3/2}]_{3/2}$ & Blend & 5.48 $\pm$ 1.10 & 6.07  & 5.90  & 5.55  & 0.60  & 0.55  & 5.03  & 4.62  & 0.85  & 0.85 \\
			          & $[(1s 2p_{1/2})_{0} 2p_{3/2}]_{3/2}$ &       &       &       &       &       &       &       &       &       &       &  \\
			He$_6$ & $[1s (2p_{3/2}^2)_2]_{5/2}$ & 51.064(6) & 2.00 $\pm$ 0.40 & 2.27  & 1.78  & 1.89  & 0.50  & 0.50  & 1.89  & 1.91  & 0.55  & 0.68 \\
			Li$_1$ & $[1s 2s^2 2p_{1/2}]_{1}$ & 46.686(5) & 2.31 $\pm$ 0.11 & 3.77  & 2.80  & 2.85  & 0.94  & 0.15  & 3.68  & 3.48  & 0.83  & 0.88 \\
			Li$_5$ & $[((1s 2s)_{1} 2p_{1/2})_{3/2} 2p_{3/2}]_{3}$ & 48.970(5) & 1.49 $\pm$ 0.14 & 2.10  & 2.14  & 1.82  & 0.44  & 0.44  & 2.02  & 2.08  & 0.56  & 0.69 \\
			Li$_6$ & $[(1s 2s)_{1} (2p_{3/2}^2)_{2}]_{3}$ & 51.154(5) & 1.11 $\pm$ 0.10 & 1.31  & 1.48  & 1.13  & 0.44  & 0.44  & 1.87  & 1.89  & 0.55  & 0.68 \\
			
			{Be$_1$} & $[1s 2s^2 2p_{1/2}^2]_{1/2}$ & 47.135(5) & 0.87 $\pm$ 0.06 & {0.58}  & 0.32  & 0.67  & 0.00  & 0.00  & 1.93  & 2.04  & 0.64  & 0.66 \\
			Be$_3$ & $[(1s 2s^2 2p_{1/2})_{0} 2p_{3/2}]_{3/2}$ & 49.349(6) & 1.75 $\pm$ 0.12 & 2.03  & 2.11  & 1.82  & 0.60  & 0.44  & 1.77  & 1.86  & 0.63  & 0.65 \\
			{Be$_4$} & $[(1s 2s^2 2p_{1/2})_{1} 2p_{3/2}]_{5/2,3/2}$ & 49.265(17) & 3.67 $\pm$ 0.32 & 3.60  & 3.77  & {3.43}  & 0.50  & 0.50  & 1.99  & 2.03  & 0.56  & 0.69 \\
			{Be$_5$} & $[1s 2s^2 (2p_{3/2}^{2})_{2}]_{5/2}$ & 51.433(6) & 2.29 $\pm$ 0.08 & 2.02  & 2.47  & {2.31}  & 0.50  & 0.47  & 1.83  & 1.85  & 0.55  & 0.68 \\
			                         & $[(1s 2s)_0 (2p_{3/2}^2)_2]_2$ &  &   &    &    &    &    &    &    &    &    &   \\
			B$_{23}$ & $[1s 2s^2 2p_{1/2}^2 2p_{3/2}]_{2}$ & Blend & 3.04 $\pm$ 0.14 & 2.75  & --  & 2.68  & 0.06 & 0.06  & 1.92  & 2.00  & 0.67  & 0.69 \\
			         & $[1s 2s^2 2p_{1/2}^2 2p_{3/2}]_{1}$ &       &       &       &       &       &       &       &       &       &       &  \\
			B$_4$ & $[(1s 2s^2 2p_{1/2})_{1} (2p_{3/2}^2)_{2}]_{3}$ & 51.603(8) & 0.89 $\pm$ 0.02 & 0.76  & 0.83  & 0.96  & 0.44  & 0.44  & 1.77  & 1.82  & 0.66  & 0.68 \\
			\hline \hline
		\end{tabular}%
	\end{table*}

	Using the data analysis procedure which combines theoretical analysis of a well-understood process (RR) into ions with different charge states with experimental
	input from the two broad band structures in Fig.~\ref{scatter}, we could finally normalize the DR resonances to the RR cross sections for each individual DR resonance peaks. 
	According to Eq.~\eqref{normalize}, the theoretical factors such as $P^{\mathrm{DR}}$, $P^{\mathrm{RR}}$, and $\sigma^{\mathrm{RR}}$ are required for the determination
	of experimental resonance strengths. These factors are calculated using three different approaches: the multiconfiguration Dirac-Fock theory (we denote by MCDF$_s$ the
	results of Ref.~\cite{Sco03} and by MCDF$_m$ the results of this work) and using the Flexible Atomic Code (FACv1.1.3)~\cite{gu2008} (results of this work). 
	Recently, the linear polarization of DR x rays $P^{\mathrm{DR}}$ was measured and benchmarked the FAC polarization predictions~\cite{Shah15}. 
	Here, we follow the the theoretical description given in Ref.~\cite{Shah15,Amaro17} to calculate the DR x-ray polarization using the FAC code. 
	The RR cross sections $\sigma^{\mathrm{RR}}$ into $n$=2 state and linear polarization of RR x~rays  $P^{\mathrm{RR}}$ are calculated according to Ref.~\cite{Sco89,gu2008}. 
	Note that, in a \emph{KLL}-DR process, there are several energetically close final states available for an intermediate state to decay into. This is due to the different
	fine-structure components occupied by the excited electrons. These transitions are characterized by different values of the degree of linear polarization.
	Hence, the $P^{\mathrm{DR}}$ represents the intensity-weighted average of polarization of those multiple final states. Since all parameters in Eq.~\eqref{normalize}
	are known now, we can determine the experimental resonance strengths and its uncertainties for each DR channel, as summarized in the fourth column in
	Table~\ref{theory-exp}, together with the observed DR resonance energies~\cite{Gonzalez06} in the third column.
	
\begin{figure}
		\begin{center}
			\includegraphics[width=0.95\columnwidth]{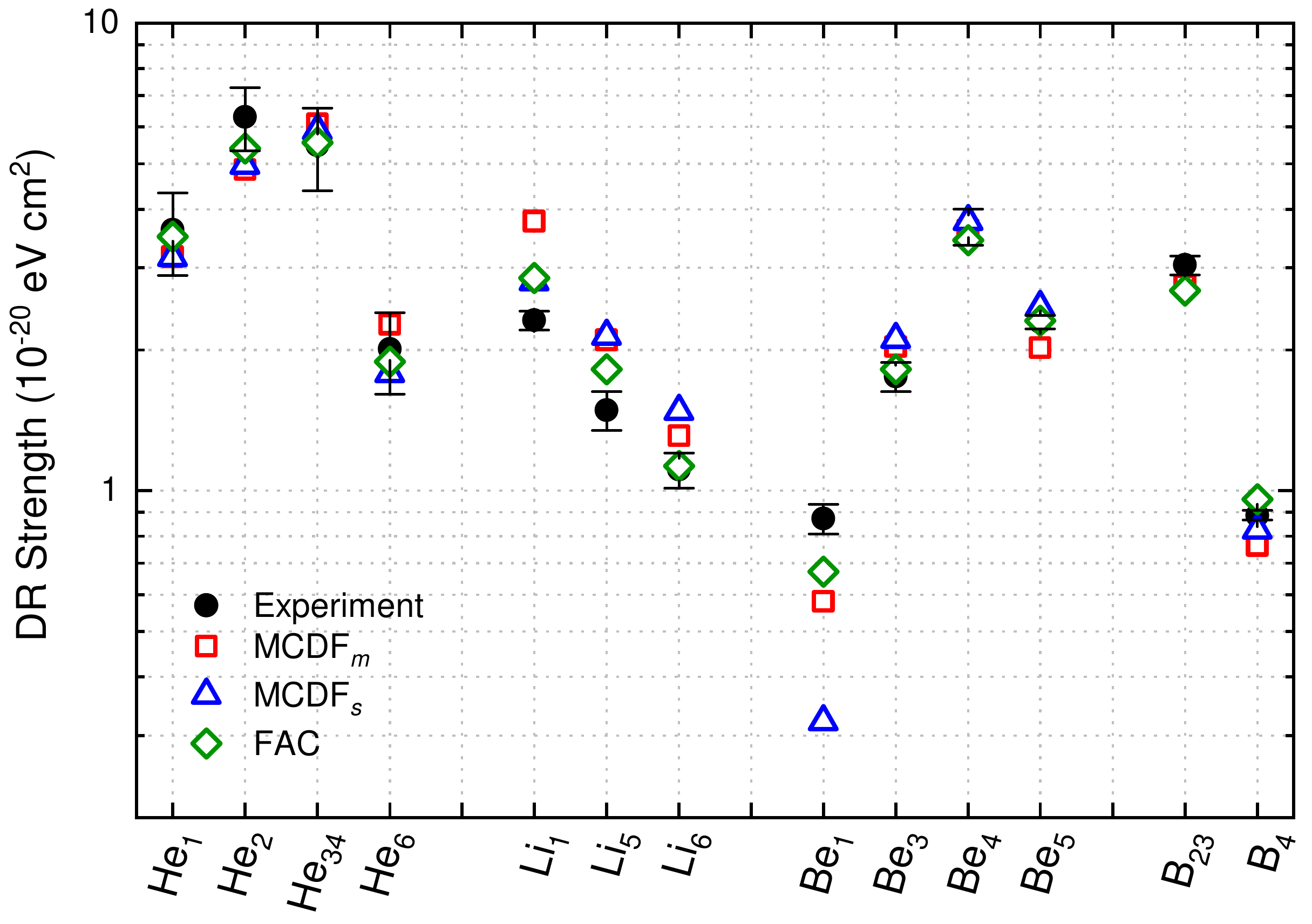}
		\end{center}
		\caption{(Color online) Comparison of experimental (solid circles) and theoretical DR strengths from MDCF$_m$ with open squares, from MDCF$_s$
			with open triangles, and FAC with open diamonds. The labeling of the resonances is explained in Table~\ref{theory-exp}.}
		\label{results}
	\end{figure}

	In Table~\ref{theory-exp}, we also compare the experimental results of resonance strengths with three theoretical calculations obtained through the MCDF and FAC methods,
	taking into account relativistic Breit interactions terms~\cite{ZH06}. Fig.~\ref{results} compares graphically the experimental results (solid circles) and the three
	calculations (open squares for MCDF$_m$, open triangles for MCDF$_s$, and open diamond for FAC results). We observe that the He-like data show a very good agreement with
	all the calculations. All the observed DR resonance strengths due to Li-like ions are slightly lower than the predictions. The FAC calculations appear closer to experimental
	values compared to MCDF values. Here, the Li$_{6}$ resonance shows good agreement with FAC prediction.
	
	The Be-like resonance strengths, in general, appear slightly scattered around the theoretical values. {For the Be$_1$ resonance, we found that it is essential
	to include the mixing of initial-state ionic configurations. In each initial state of DR, the total electronic wave function is described by the ionic ground state,
	complemented with the corresponding partial wave of the incoming continuum-state electron, as implied in Eq.~(\ref{eq:dr-rate}).
	Specifically, in case of the Be$_1$ line, the mixing of the $1s^2 2s^2$ and $1s^2 2p_{1/2}^2$ configurations is relevant, as the latter has an almost identical orbital occupation
	as the Be$_1$ $[1s 2s^2 2p_{1/2}^2]_{1/2}$ autoionizing state, thus they largely overlap in space and yield a sizable capture matrix element. The MCDF$_m$ and FAC calculations account for this effect, while MCDF$_s$ does not.
	Other resonances and charge states were found to be not affected by such initial-state mixing effects.
	The Be$_3$ line shows the best agreement with the FAC prediction, while the Be$_4$ and Be$_5$ resonances agree with both FAC and MCDF results. We did not find a particular
	reason for the difference between FAC and MCDF for Be$_3$ line.} For B-like resonances, both MCDF and FAC predictions agree with the experimental strengths.
	
	In all cases, the agreement between theoretical and experimental resonance strengths can be regarded as satisfactory, given the complexity of the autoionizing states
	involved. Furthermore, as the strength of a resonance as observed by detecting the emitted x rays depends on the angular distribution of the radiation emission, such
	measurements are more sensitive to the details of the theoretical calculations than experiments where total recombination cross sections are directly determined. 
	E.g. as it was shown by Fritzsche et al.~\cite{Fritzsche08}, the mixing of the $E1$ and $M2$ multipolarities in the radiative decay process may cause an observable change
	in the angular differential cross sections for high-\textit{Z} ions. Moreover, the influence of electron interaction corrections due to magnetic and
	retardation effects (i.e. the Breit interaction) was shown to modify the linear polarization of DR x rays as well as the resonance strengths~\cite{Fritzsche09,Shah15,Amaro17}. 
	{Note that the present experiment was performed using a mixture of naturally abundant Hg isotopes. It contains $^{199}$Hg (17 \%) and $^{201}$Hg (13 \%) with
	nuclear spins 1/2 and 3/2, respectively. The hyperfine interaction may reduce the resulting anisotropy of DR x rays, as it was shown in Refs.~\cite{Surz13,Wu14,Wu16,Zaytsev17},
	and its inclusion in the theoretical description of resonance strengths could potentially improve the agreement with the experiment.}
	%

	%
	%
	\section{Scaling formulae}\label{sect:scaling}
	%
	%
	\subsection{Total~\emph{KLL} DR strength}\label{sect:total}
	%
	%
	
	The total DR resonance strength for He-like Hg ions can be summed up over all levels and charge states (see Table~\ref{theory-exp}), and is found to be
	$(20.4 \pm 1.9) \times 10^{-20}$ eV cm$^2$ which can be favorably compared with the theoretical values of 20.3 (MCDF$_m$), 19.7 (MCDF$_s$), and
	22.2 ({FAC}) $\times$ 10$^{-20}$ eV cm$^2$.
	
	In previous years, the total~\emph{KLL} resonance strengths of He-like ions have been measured by a number of experiments in various low- and mid-$Z$
	ions~\cite{Kilgus93,Ali11,Ali91,ORourke04,Beiersd92,Knapp89,Zhang04,Fuchs98,Knapp93,Watanabe07,Yao10}, while data for very heavy ions, where the relativistic and QED
	effects play a critical role are still scarce~\cite{Kavanagh2010,Tu2016}. By using the results of the present experiment along with previously
	reported measurements, we can shed light on the tendency of the strength as a function of the nuclear charge number and provide information on its behavior
	at the upper end of the curve. 
	
	It is known that most of the quantities describing the DR resonance strength in Eq.~\eqref{eq1} have clear dependence on the atomic number $Z$. 
	In a completely nonrelativistic formalism, the DR resonance strengths are expected to be proportional to $Z^2$ at low $Z$. This is due to the fact that the autoionization
	rate $A_a^d$ is roughly independent of $Z$, the radiative transition rate $A_r^d$ scales as $Z^4$~\cite{bambynek}, and  the DR resonance energy $E_{\mathrm{DR}}$
	is approximately proportional to $Z^2$.  Therefore, using Eq.~\eqref{eq1}, the $Z$-dependence of the DR resonance strength $S^{\mathrm{DR}}$ can be described as follows:
	\begin{eqnarray}\label{oldscaling}
	S^{\mathrm{DR}} \propto \frac{1}{Z^2} \frac{Z^4 Z^0}{m_1 Z^4 + m_2 Z^0}  = \frac{1}{m_1 Z^2 + m_2 Z^{-2}}\;,
	\label{eq3} 
	\end{eqnarray}
	where $m_1$ and $m_2$ are fit parameters and can be calculated, in a first nonrelativistic approximation, from nonrelativistic hydrogenic wave functions~\cite{Watanabe01}. 
	In a similar way, beyond first-order dielectronic recombination, the $Z$-scaling laws for trielectronic and quadruelectronic recombination were also derived,
	see Eqs.~(9) and (10) of Ref.~\cite{Beilmann13}. 
	
\begin{figure}
		\begin{center}
			\includegraphics[width=0.95\columnwidth]{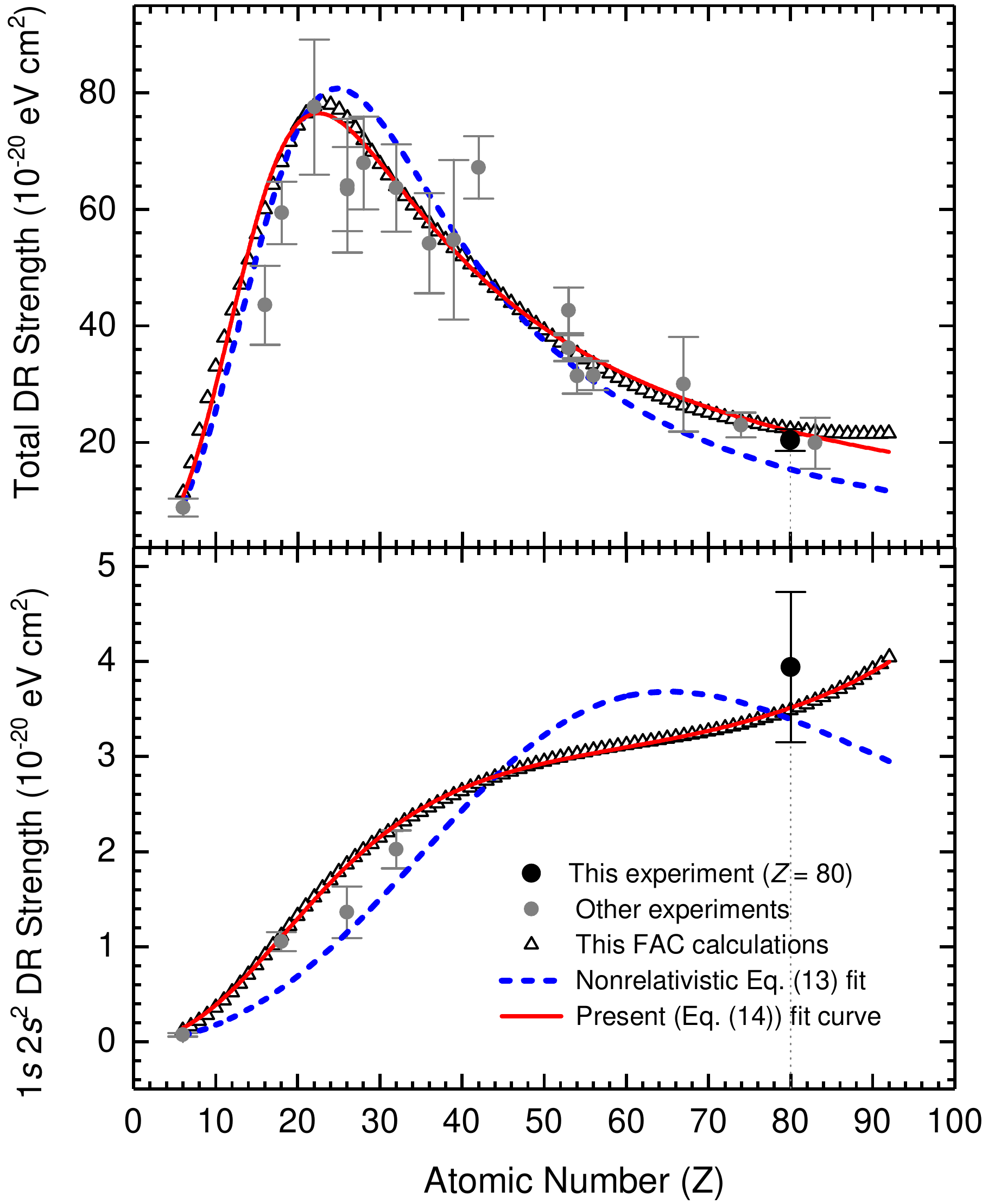}
		\end{center}
		\caption{(Color online) 
			Observed total (top) and partial (bottom) \emph{KLL} DR resonance strengths for He-like ions as a function of the atomic number $Z$. 
			The stars with vertical dashed line represent the experimental results of Hg$^{78+}$ ions. 
			The other data in solid circle are C$^{4+}$~\cite{Kilgus93}, S$^{14+}$~\cite{Ali11}, Ar$^{16+}$~\cite{Ali91}, Ti$^{20+}$~\cite{ORourke04}, Fe$^{24+}$~\cite{Beiersd92,Kavanagh2010}, Ni$^{26+}$~\cite{Knapp89}, Ge$^{30+}$~\cite{Zhang04}, Kr$^{34+}$~\cite{Fuchs98}, Y$^{37+}$~\cite{Kavanagh2010}, Mo$^{40+}$~\cite{Knapp93}, I$^{51+}$~\cite{Watanabe07,Kavanagh2010}, Xe$^{52+}$~\cite{Yao10}, Ba$^{54+}$~\cite{Knapp93}, Ho$^{65+}$~\cite{Kavanagh2010}, W$^{72+}$~\cite{Tu2016}, and Bi$^{81+}$~\cite{Kavanagh2010}. 
			The dashed blue curve represents the Eq.~\eqref{eq3} fit to the FAC data (open triangles), whereas the best-fitted DR strengths according to Eq.~\eqref{newscalingnew} is shown by a solid red curve. The fit parameters are represented in Table~\ref{tab:fitpars}.
		}
		\label{results2}
	\end{figure}

	The top panel of the Fig.~\ref{results2} shows the result of the present experiment and all previous experimental results of total DR resonance strengths for He-like
	ions as a function of atomic number. With the help of FAC code, we also calculated total DR resonance strength from $Z$ = 6 to 92 taking into account the Breit interaction
	in the calculation of the Auger rates. 
	The theoretical FAC data are shown in open triangles in Fig.~\ref{results2}. Since most of the experiments at mid- and high-$Z$ show a satisfactory agreement with FAC
	predictions and experimental data at low-$Z$ are very sparse, we determine to fit the Eq.~\eqref{eq3}~\cite{Watanabe01} to the FAC data instead of experimental data in
	order to improve the uncertainties in the parameters $m_1$ and $m_2$. The blue dashed curve in Fig.~\ref{results2} represents the fit via Eq.~\eqref{eq3}.
	The best fit parameters were found to be $m_1=(1.00 \pm 0.02) \times 10^{15}$ eV$^{-1}$ cm$^{-2}$ and $m_2=(3.81 \pm 0.11) \times 10^{20}$ eV$^{-1}$ cm$^{-2}$ with
	$\chi^2/\mathrm{d.o.f.} = 27.9$.

	In this plot, a slight deviation between the FAC and the Eq.~\eqref{eq3} fit curve can easily be noticed for the ions with higher nuclear charge. 
	The experimental values for $Z$ = 67 (Ho), 74 (W), 83 (Bi), and our present results for Hg $Z$ = 80 show likewise disagreement with the Eq.~\eqref{eq3} fit curve. 
	Such deviation can be expected since relativistic effects give large correction to the non-relativistic autoionization rates $A_a^d$~\cite{Zim90}. 
	In~Eq.~\eqref{eq3}, the leading non-relativistic autoionization term corresponds to the expression $m_2 Z^{-2}$ in the denominator. 
	We correct~Eq.~\eqref{eq3} with relative order $(\alpha Z)^2$ in order to describe the leading Breit term and a correction of relative order $(\alpha Z)^3$
	in order to take higher-order many-electron relativistic correction into account. 
	With these amendments, the following functional form appears suitable, and we would like to refer to it as a semi-empirical scaling law: 
	\begin{eqnarray}
	S^{\mathrm{DR}}= \frac{1}{m_1 Z^2 + m_2 Z + m_3 + m_4 Z^{-2}} \,.
	\label{newscalingnew}
	\end{eqnarray}
	The red curve in the top panel of Fig.~\ref{results2} show a fitting result with the use of Eq.~\eqref{newscalingnew} and the best fitting parameters are given Table~\ref{tab:fitpars}.
	It can easily be observed that the new semi-empirical formula fits the FAC data exceptionally well compared to the Eq.~\eqref{eq3}. 
	Moreover, it also improves the $\chi^2/\mathrm{d.o.f.}$ value from 27.9 to 2.1.

\begin{table*}
	\centering
	\caption{The parameters obtained by fitting Eq.~\eqref{newscalingnew} to both total and partial ($1s 2s^2$) resonance strengths data obtained by FAC. The uncertainties here are given as 1$\sigma$.}
	\begin{tabular}{lcccc}
		\hline \hline
		& $m_1$ ($\times 10^{15}$ eV$^{-1}$ cm$^{-2}$) & $m_2$ ($\times 10^{16}$ eV$^{-1}$ cm$^{-2}$) & $m_3$  ($\times 10^{17}$ eV$^{-1}$ cm$^{-2}$) & $m_4$  ($\times 10^{20}$ eV$^{-1}$ cm$^{-2}$) \\
		\hline
		Total resonance strengths & $0.11 \pm 0.04 $ & $5.62 \pm 0.35$ & $-7.00 \pm 0.81 $ & $3.47 \pm 0.09$  \\
		$1s 2s^2$ resonance strengths & $-5.30 \pm 0.15 $ & $70.5 \pm 2.19$ & $20.55 \pm 8.47 $ & $252.67 \pm 2.93$ \\
		\hline \hline
	\end{tabular}%
	\label{tab:fitpars}%
\end{table*}%

	%
	\subsection{The $\boldsymbol{1s 2s^2}$ DR resonance}\label{sect:partial}
	%
	%

	The particular DR channel via the $1s 2s^2$ state is interesting because the radiative decay of this autoionizing state preferably proceeds via electric dipole ($E$1)
	transition involving simultaneous two-electron decay, forming a final $1s^2 2p$ state while emitting a single x-ray photon (see, e.g. Ref.~\cite{Dong}). 
	As its DR strength is expected to be small in low-$Z$ ions, only a few experimental observations were reported so far~\cite{Mannervik97,Beiersd92,Zou03,Zhang04}. 
	The observed partial DR strengths including the present data for Hg are plotted in the bottom panel of Fig.~\ref{results2}.  It is easily found that the partial
	strengths for low-$Z$ ions are indeed very small (less than one percent of the total DR strength) but, in Hg ions, the partial DR strength for this state - labeled
	as He$_1$ in Table~\ref{theory-exp} -- reaches nearly 20 \% of the total DR strengths.

	The top and bottom panel of Fig.~\ref{results2} shows that the total and partial DR strengths reach maximum at very different nuclear charges. It can be understood as follows: 
	According to recent calculations~\cite{Dong}, the radiative rates from this state in low-$Z$ ions increase as $Z^4$ but are still orders of magnitude smaller than the
	autoionization rates which are nearly independent of the nuclear charge number of the ion. It should also be noted that, although higher-order transitions, in particular,
	magnetic dipole ($M1$) transitions increase proportionally to $Z^{10}$, their transition rates are still too small to significantly influence the overall transition rates
	of this particular state. Thus, as expected from Eq.~\eqref{eq1}, a few observed data of the partial DR strength shown in Fig.~\ref{results2} seem to follow such a
	$\sim Z^2$ scaling in the low-$Z$ regime, similarly to the total DR strength shown in Eqs.~\eqref{oldscaling} and \eqref{newscalingnew}. However, the observed partial
	strength data for high-$Z$, though deviating from the $\sim Z^2$-dependence, still increase roughly as $Z^1$ with increasing $Z$. This feature is in a sharp contrast
	to that observed in total DR strengths which decrease roughly as $Z^{-2}$ in the high-$Z$ region. This can be explained in following way: although for very heavy ions,
	the autoionization and radiative rates increase as $Z^2$ and $Z^4$, respectively, both rates become comparable and the total transition rates (in the denominator of
	Eq.~\eqref{eq1}) increase, on average, roughly as $Z^3$ in the very high-$Z$ ion regime. Thus, following Eq.~\eqref{eq1}, it is found that the partial DR strengths for
	this particular state increase as $Z^1$, agreeing with those observed and shown with the red solid curve in the bottom panel of Fig.~\ref{results2}.
	
	As the experimental data for the partial DR strength for this particular state are too scarce, we cannot provide any definite conclusion in regard to the present scaling law. 
	Therefore, we use again Eq.~(\ref{newscalingnew}) to fit the theoretical FAC data and the parameters obtained by fitting are given in Table~\ref{tab:fitpars}. 
	By comparing the fits of Eq.~\eqref{oldscaling} (blue dashed curve) and Eq.~\eqref{newscalingnew} (red solid curve) in the bottom panel of Fig.~\ref{results2}, one can see
	that the new scaling law gives a considerably better fit even for the state-resolved resonance strength of the $1s2s^2$ state.

	%
	%
	\section{Summary}
	\label{sect:summary}
	%
	%
	
	In the present work, we have determined the~\emph{KLL} DR resonance strengths for charge- and electronic-state-specific highly charged mercury ions,
	ranging from the He-like to the B-like charge state through observing x~rays emitted both from the DR and RR processes.
	Our work leads to a pathway of determining~\emph{KLL} DR resonance strengths in an absolute normalization and allowed us to gain new insights into a dynamical
	aspect of processes in an EBIT driven at high fields. 
	The measured DR resonance strengths were compared with two different atomic structure methods, MCDF and FAC. The effect of the Breit interaction, a relativistic retardation and
	magnetic correction to the electron-electron interaction, was included in the dielectronic capture matrix elements. Theoretical results have been found to be generally in good
	agreement with the experimental data, except for some resonances, given in Table~\ref{theory-exp}. The reason for the discrepancies is unknown at present.
	
	The present work also sheds light to the tendency of the resonance strength ${S}^{\mathrm{DR}}$ as a function of the atomic number, especially to the behavior of the
	resonance strengths in the high-$Z$ regime. We present a compact $Z$-scaling formula for both the total and partial~\emph{KLL} DR strengths as a function
	of the atomic number $Z$ of the ions involved. The difference in the $Z$-scaling between the total (integrated) and partial ($1s 2s^2$ state in initially He-like ions)
	resonance strengths was discussed in detail. A new semi-empirical formula, Eq.~(\ref{newscalingnew}), improves the non-relativistic $Z$-scaling formula~\cite{Watanabe01} by including
	relativistic corrections, thus extending the range of applicability to the high-$Z$ domain. Such an improved $Z$-scaling law for DR strengths can also be
	useful to produce large sets of atomic data needed for the modeling and diagnostics of magnetically confined fusion plasmas~\cite{beiersdorfer2015} and hot
	astrophysical plasmas~\cite{beiersdorfer2003,hitomi18}.

	\begin{acknowledgments}
		We are very thankful to Prof.~J.~H.~Scofield for providing his theoretical results and Prof.~C.~Z.~Dong for discussions about his work~\cite{Dong} on the transition rates of the $1s 2s^2$ state. {
	Also, we thank Dr. M. F. Gu and Dr. N. Hell for their help with specific features of the FAC code.} The work of U.D.J. was supported by the National Science Foundation (Grant PHY-1710856).
	\end{acknowledgments}
	Z.H. and C.S. contributed equally to this work.

%

\end{document}